\documentclass[12pt]{article}

\usepackage{amsmath,amssymb}
\usepackage{graphicx}
\usepackage{caption}
\usepackage{color}
\newcommand{\redc}[1]{{\color{black} #1}} 
\usepackage{dcolumn}
\usepackage{bm}
\usepackage[numbers,super,comma,sort&compress]{natbib}

\definecolor{background-color}{gray}{0.98}
\title{Shannon Entropy and Many-Electron Correlations: Theoretical Concepts, Numerical Results and Collins Conjecture\footnote{Invited paper on International Journal of Quantum Chemistry for the special issue ``Quantum Information in Chemistry'' (2015)}}
\author{Luigi Delle Site \thanks{Institute for Mathematics, Freie Universit\"{a}t Berlin, Berlin, Germany; e-mail: luigi.dellesite@fu-berlin.de}}

\begin{document}
\maketitle
\begin{abstract}
In this paper I will discuss the overlap between the concept of Shannon Entropy and the concept of electronic correlation. Quantum Monte Carlo numerical results for the uniform electron gas are also presented; these latter on the one hand enhance the hypothesis of a direct link between the two concepts but on the other hand leave a series of open questions which may be employed to trace a roadmap for the future research in the field. 
\end{abstract}

\clearpage


  \makeatletter
  \renewcommand\@biblabel[1]{#1.}
  \makeatother

\bibliographystyle{apsrev}

\renewcommand{\baselinestretch}{1.5}
\normalsize

\clearpage

\section*{\sffamily \Large INTRODUCTION} 
In a seminal paper\cite{shannon} Claude Shannon formalized the idea of assigning probabilities to the outcome of uncertain events and introduced the entropy as measure of uncertainty. Nowadays the concept of Shannon entropy has crossed a considerable number of barriers between traditionally separated disciplines and became a universal concept of statistical physics. In particular in molecular physics its use is spreading in several directions, from modeling hydrophobic interactions (see e.g.Ref.\cite{hummer}), to simulation of rare events (see e.g. Refs.\cite{perilla,suarez}), to the treatment of chemical bonds and electronic properties (see e.g. Refs.\cite{romanbook,liu}), to cite a few. Given the universality claimed above, it became obvious that the Shannon entropy would come in contact with another major concept of modern physics, that is  electronic correlations.
The definition and calculation of electronic correlations represents nowadays the Holy Grail of electronic structure calculations in condensed matter, material science, chemical physics and biochemistry/physics\cite{luca}. The fast development of modern technology requires the treatment of large systems at an accuracy that goes much beyond the accuracy offered by current computational approaches.
Electronic correlations play, in this sense, a key role; approaches which can properly account for electronic correlations as Quantum Monte Carlo (QMC) and high level quantum chemical methods, are computationally too expensive and thus restricted to small systems, instead Density Functional Theory (DFT), that is the most popular electronic structure method, can treat relatively large systems at manageable computational prices but has got in the electronic correlations its most empirical aspect. The empiricism of DFT regarding the description of electronic correlations is translated either into poor accuracy or into designing correlations functionals whose physical justification is not clear. These functionals are often numerically successful but, being poorly understood regarding their physical meaning, are not universal or at least not sufficiently transferable\cite{savin}. The discussion above leads on the one hand to the question of how to improve the computational performance of the more accurate computational approaches and on the other hand underlines the need to investigate new directions along which the electronic correlations can be better described within the DFT approach. This paper treats the second case by exploring one possible direction, that is the hypothesis of employing the concept of Shannon entropy into the development of electronic correlation functionals. I will report recent ideas where the Shannon entropy has been linked to DFT and in particular to the concept of electronic correlations. Moreover I will show recent QMC results for a gas of electrons which, within certain limitations, support the link between electronic correlations energies and Shannon Entropy. Possible connections between the electron density of DFT and the many-body wavefunctions/correlations are put forward in terms of encoding/decoding concepts of information theory. 
However, despite the intriguing results, a critical appraisal is mandatory and the point of view of the skeptical  part of the community needs to be reported as well. The final conclusions are a series of open questions which, if properly addressed, can certainly confirm whether or not the concept of Shannon entropy is useful to push forward the understanding of electronic correlations and can lead to a practical implementation within DFT based computational approaches.  
\section*{\sffamily \Large CONCEPTS}
In this section I will briefly review the general features of the two main concepts of this paper, namely the Shannon entropy and the electronic correlations.
The description of each concept is done with an emphasis on the aspects which are of relevance for the later discussion.
\subsection*{\sffamily \large Shannon Entropy}
{\it Definition:} For a given probability distribution $p(x)$, the Shannon entropy is defined as:
\begin{equation}
S=-\sum_{x}p(x)\log_{2}p(x)
\end{equation}
with $x$ being a discrete variable indicating specific outcomes in a set of events, or:
\begin{equation}
S=-\int p(x)\log_{2}p(x)dx
\label{esse}
\end{equation}
if $x$ is a continuous variable. The logarithm base $2$ is due to the fact that the idea was originally developed within the framework of binary language, however it can be extended to any  base, being the conversion simply a multiplicative constant.
The definition above is based on the {\bf Shannon Source Coding Theorem} \cite{shannon}, elaborated by Shannon as an answer to the following question \cite{petz}:\\
{\it Suppose we have a set of possible events whose probabilities of occurrence are $p_{1}$, $p_{2}$....$p_{n}$. These probabilities are known but that is all we know concerning which event will occur. Can we find a measure of how much ``choice'' is involved in the selection of the event or how uncertain we are of the outcome?}\\
Thus, Shannon Entropy is a measure of the uncertainty or in other terms gives an absolute limit on the best possible average length of encoding of information (for example the length of a sequence/combination of $0$'s and $1$'s in a binary code) without loss, if the information can be represented by a probability distribution. The description above is the essence of the concept needed for my discussion later on.
\subsection*{\sffamily \large Correlation Energies in Many-Electron systems}
I will define the ground state correlation energy of a $N$-electron system as the difference between the exact kinetic and electron-electron energy of ground state and the equivalent Hartree-Fock energy of the system:
\begin{equation}
E_{c}=\left<\psi_{ex}|\widehat{H}_{ee}|\psi_{ex}\right>-\left<\phi_{HF}|\widehat{H}_{ee}|\phi_{HF}\right>
\label{corrdef}
\end{equation}
With $\widehat{H}_{ee}=\widehat{T}+\widehat{V}_{ee}$, that is the sum of the one-particle kinetic operator and of the two-particle Coulomb electron-electron operator. The wavefunctions $\psi_{ex}({\bf r}_{1},{\bf r}_{2},.....{\bf r}_{N})$ and $\phi_{HF}({\bf r}_{1},{\bf r}_{2},.....{\bf r}_{N})$ are respectively, the exact solution and the single Slater determinant approximation of the Schr\"{o}dinger problem: $H\psi=\epsilon \psi$, for the ground state with Hamiltonian: $\widehat{H}=\widehat{T}+\widehat{V}_{ee}+\widehat{V}_{ext}$, with $+\widehat{V}_{ext}$ the external potential (e.g. electron-nucleus Coulomb interaction).
\redc{At this point a clarification is needed regarding the general concept of ``{\it electronic correlations}''. The definition I have chosen is usually indicated as ``{\it Quantum Chemical}'' electronic correlation \cite{bookdft}, as defined by  P.O.L\"{o}wdin \cite{low}. This definition is the most natural if one considers wavefunctions explicitly, as it is traditionally done by quantum chemists; however a certain amount of electron correlation is already described by the Hartree-Fock approximation, that is the electron exchange term which describes the correlation between electrons with parallel spin. In this paper I will not consider the exchange term because for the case I will explicitly treat (uniform gas) analytic formulas for the exchange energy are known. A further clarification is needed; I will embed the discussion within the framework of DFT, the definition I have chosen would not correspond to the definition of correlation energy of Kohn-Sham DFT approach. The difference between L\"{o}wdin and Kohn-Sham definition is (see Ref.\cite{bookdft} for details):
\begin{equation}
E_{c}^{KS}-E_{c}=E^{HF}-E_{x}^{KS}[\rho]
\label{kscorr}
\end{equation}
where $E_{x}^{KS}[\rho]$ is the exchange energy written in terms of Kohn-Sham orbitals (for simplicity here spin variables are not explicitly considered):
\begin{equation}
E_{x}^{KS}[\rho]=-\frac{1}{2}\sum_{kl}\Theta_{k}\Theta_{l}\int d{\bf r}\int d{\bf r}^{'}\phi^{*}_{k}({\bf r})\phi_{l}({\bf r})\frac{1}{|{\bf r}-{\bf r}^{'}|}\phi^{*}_{l}({\bf r}^{'})\phi_{k}({\bf r}^{'})
\end{equation}
where $\Theta_{i}$ is the occupation number of the Kohn-Sham orbital $\phi_{i}({\bf r})$. My choice of definition of electronic correlation, when one goes beyond the uniform electron gas, does not apply directly to Kohn-Sham DFT. If one wants to apply the concepts of this paper to  Kohn-Sham DFT the discussion above must be taken into account and the term $E_{x}^{KS}[\rho]$ should be properly treated. However  L\"{o}wdin definition is general enough and it applies directly to the calculation of energy functionals (above all the kinetic functional) in DFT when  approaches other than Kohn-Sham are employed (e.g. Orbital Free DFT\cite{trickeyl}). It may seem reductive, since the Kohn-Sham approach, because of its success, is often identified with DFT itself; moreover, the use of orbitals allows a sort of Hartree-Fock-like analysis in terms of individual electrons to chemists and allows solid state physicists to think in terms of band structures, however, in my view, to go beyond Kohn-Sham orbitals is a direction which shall be supported. Its potentiality  has never really been explored in full and it may turn extremely useful in connection with other electronic structure methods (see e.g. \cite{ijqc}) or in modern popular multiscale studies (see e.g. \cite{bern}).} 
In any case, given the definition chosen, for later discussions let us define specific quantities:
\begin{equation}
T_{c}=\left<\psi_{ex}\left|\widehat{T}\right|\psi_{ex}\right>-\left<\phi_{HF}\left|\widehat{T}\right|\phi_{HF}\right> 
\label{tcorr}
\end{equation}
that is the correlation part of the kinetic energy, and:
\begin{equation}
V_{c}=\left<\psi_{ex}\left|\widehat{V}_{ee}\right|\psi_{ex}\right>-\left<\phi_{HF}\left|\widehat{V}_{ee}\right|\phi_{HF}\right>.
\label{vcorr}
\end{equation}
Moreover I define the kinetic correlation energy density as:
\begin{equation}
t_{c}({\bf r})=\left[\frac{1}{\rho_{ex}({\bf r})}\int\psi^{*}_{ex}\widehat{T}\psi_{ex}d{\bf r}_{2}.....d{\bf r}_{N}-\frac{1}{\rho_{HF}({\bf r})}\int\phi^{*}_{HF}\widehat{T}\phi_{HF}d{\bf r}_{2}.....d{\bf r}_{N}\right]
\label{tdens}
\end{equation}
and the potential correlation energy density as:
\begin{equation}
v_{c}({\bf r})=\left[\frac{1}{\rho_{ex}({\bf r})}\int\psi^{*}_{ex}\widehat{V}_{ee}\psi_{ex}d{\bf r}_{2}.....d{\bf r}_{N}-\frac{1}{\rho_{HF}({\bf r})}\int\phi^{*}_{HF}\widehat{V}_{ee}\phi_{HF}d{\bf r}_{2}.....d{\bf r}_{N}.\right]
\label{vdens}
\end{equation}
where $\rho_{ex}({\bf r})=\int|\psi_{ex}|^{2} d{\bf r}_{2}...d{\bf r}_{N}$, with $\int\rho_{ex}({\bf r}) d{\bf r}=N$ and $\rho_{HF}({\bf r})=\int|\phi_{HF}|^{2}d{\bf r}_{2}...d{\bf r}_{N}$, with $\int\rho_{HF}({\bf r}) d{\bf r}=N$. 
The two definitions above represent the key quantities of the correlation energy functional in DFT.\\
Strictly speaking the correlation energy is traditionally defined as the difference between the total energy of the exact ground state and the total energy of the corresponding Hartree-Fock solution, although it has been underlined in Ref.\cite{vignale} that this definition is not the unique one. For this discussion the  definition of energy is not crucial, but it is crucial the definition of energy density. The definition of energy density that I have adopted here is consistent with both, my definition of correlation energy and the definition usually adopted in literature. In fact for the exact ground state and the for Hartree-Fock solution, the energy density of the external potential is the same, while this is not true for total energies due to the possibility that $\rho_{ex}$ and $\rho_{HF}$ are different. 
\subsection*{\sffamily \large Recurrence of Shannon Entropy into the description of Many-Electron Systems}
The concept of Shannon entropy is starting to become a powerful tool for the description of many-electron systems. One example is the work of A.Nagy  who has shown, in a rigorous way, the property of the Shannon entropy density as descriptor of Coulomb system \cite{nagy1}. The Shannon entropy in this case is defined as:
\begin{equation}
S=-\int\rho({\bf r})\log\rho({\bf r})d{\bf r}
\label{shentrho}
\end{equation}
and the Shannon entropy density:
\begin{equation}
s({\bf r})=-\rho({\bf r})\log\rho({\bf r})
\end{equation}
with $\rho({\bf r})$, being the three dimensional electron density, as previously defined.
The results of Ref.\cite{nagy1} are very interesting because they may offer an alternative approach (and thus perspective) to the description of electronic systems. Moreover, in collaboration with S.Liu, Nagy has proposed the interpretation of the gradient of Shannon entropy density per particle, $-\log\rho({\bf r})$, as local wave-vector\cite{liunagy}.
This result is potentially very interesting since it makes a link between a quantum object directly connected to the system's wavefunction (wave-vector) and a statistical measure of uncertainty of the corresponding electron distribution (Shannon entropy). \redc{Other applications concern, for example, the use of Shannon Entropy as indicator of avoided crossings in atomic spectroscopy for electronic systems in the presence of magnetic and electric fields \cite{uno} or the study of relevant chemical reactions like biomolecular nucleophilic substitutions reactions \cite{due,tre,quattro, sette}}.
Moving forward, of particular interest for the discussion in this paper is the recurrence of the idea of logarithm (and directly or indirectly of the Shannon entropy) to describe electronic correlations.
The use of the logarithm of a distribution, and thus something strictly related to the Shannon Entropy, as a statistical measure of the correlation strength was put forward, for example, by Gottlieb and Mauser \cite{gottmaus}. They quantify the electronic correlation in a wavefunction $\psi$ by comparing the wavefunction to the uncorrelated state, $\Gamma$, which has the same one particle statistical operator, $\gamma$, of $|\psi><\psi|$. The definition of $\gamma$ is such that: $\gamma({\bf r},{\bf r'}) =N\int \psi({\bf r},{\bf r}_{2}.....{\bf r}_{N})\psi({\bf r'},{\bf r}_{2}.....{\bf r}_{N}) d{\bf r}_{2}...d{\bf r}_{N}$.
Where $({\bf r}_{2}....{\bf r}_{N})$ are the coordinates of the $N-1$ particles over which the integration is performed and ${\bf r}$ and ${\bf r'}$ refer to the electron chosen as reference.
The strength of correlation is then defined as:
\begin{equation}
Corr(\psi)=-\log\left<\psi,\Gamma\psi\right>
\end{equation}
that is the logarithm of the dissimilarity between $\psi$ and its projection on $\Gamma$. Along the same lines, but more closely to the idea of Shannon entropy is for example the recent work of Byczuck {\it et al.} \cite{bycz}. They use the concept of von Neumann entropy \cite{vonn}, to define a measure of correlation, by calculating the relative entropy of a quantum state with respect to an uncorrelated product state.
The von Neumann entropy is defined as: $S_{vn}=-tr(\widehat{\rho}\log\widehat{\rho})$ where $\widehat{\rho}=\sum_{i}p_{i}|\psi_{i}><\psi_{i}|$ is the density operator built on the quantum states $|\psi_{i}>$ each of which has probability $p_{i}$.The  von Neumann entropy corresponds to the Shannon entropy if $|\psi_{i}>$'s are eigenstates of the system. Similar ideas start to increasingly spread in literature and the concept of statistical determination of the correlation strength is involving in an increasing way the Shannon entropy. For example, Sagar {\it et al.} \cite{sagar}, proposed the measure of Mutual Information to define the strength of correlation for spherically symmetric systems up to a two-particle correlation approximation. The concept of Mutual Information is such that the measure of correlation is defined as:
\begin{equation}
I=\int f(r_{1},r_{2})\log\frac{f(r_{1},r_{2})}{\sigma(r_{1})\sigma(r_{2})}d{\bf r}_{1}d{\bf r}_{2}
\label{sag}
\end{equation}
where $f(r_{1},r_{2})$ is the spin-traced spherically averaged two-particle pair distribution function and $\sigma(r)$ the spherically averaged one-electron density (that is $\frac{\rho({\bf r})}{N}$).  Eq.\ref{sag}  can be shown to be the difference between the one-particle and the two-particle Shannon entropy. Such a difference can be interpreted as the difference between the one particle and two-particle localization, that is a measure of the pairing of the particles, in this sense it can be interpreted as a measure of electronic correlation. Narrowing down to interpretations of Shannon entropy which can be related to the DFT or related theories, I have found interesting 
the idea of defining the measure of correlation or delocalization of the electron cloud $\rho({\bf r})$ as done by Romera and Dehesa \cite{romdeh} \redc{(see also Ref.\cite{sei})}. They define the measure of correlation as:
\begin{equation}
J_{\sigma}=\frac{1}{2\pi}e^{\frac{2}{3}S_{\sigma}}
\label{exp}
\end{equation}
where $S_{\sigma}=-\int\sigma({\bf r})\log\sigma({\bf r})d{\bf r}$.
The authors specify that $J_{\sigma}$ measures the electronic correlation because the smaller $S$ the more concentrate is the wavefunction of the state and thus the electron is more localized and local interactions dominate on long range correlations. On the contrary the larger $S$ the more delocalized is the wavefunction of the state and thus the more dominant the long range correlations.
Moreover $J_{\sigma}$ is characterized by scaling properties which give to it a solid physical consistency. The idea of localization v.s. delocalization as well as the scaling properties are important concepts in the development of energy functionals within the DFT framework and they will be of relevance in the discussion later. Inspired by the work of Romera and Dehesa I have proposed a kinetic functional of the form $T[\rho]=T_{w}[\rho]+\alpha e^{wS[\rho]}$, where $T_{w}[\rho]$ is the Weizsacker term \cite{epl}. This conclusion has been reached by calculating the kinetic energy density for a uniform gas of interacting electrons via a sampling of the electron-electron interactions using a many-body distribution \cite{prb,jcpluca}. The basic ingredients of the model used are reported below.
The procedure is based on the seminal paper of Sears, Parr and Dinur, the pioneers in linking DFT with information theory \cite{sears}. They use the factorization: $\psi({\bf r},{\bf r}_{2},....{\bf r}_{N})|^{2}=\rho({\bf r})f({\bf r}_{2}.....{\bf r}_{N}|{\bf r})$ to determine the energy functional expression.
\redc{This factorization leads to a local, non-interacting, kinetic energy term proportional to the Fisher Information (Weizsacker term): $I[\rho]=\frac{1}{8}\int\frac{|\nabla\rho({\bf r})|^{2}}{\rho({\bf r})}d{\bf r}$ (see also Ref.\cite{nagyreq})} and to a non local term: $I_{nloc}[\rho]=\frac{1}{8}\int\rho({\bf r})\left[\int\frac{|\nabla_{{\bf r}}f({\bf r}_{2}.....{\bf r}_{N}|{\bf r})|^{2}}{f({\bf r}_{2}.....{\bf r}_{N}|{\bf r})}d{\bf r}_{2}....d{\bf r}_{N}\right]d{\bf r}$. 
It must be noticed that the exact knowledge of $f$ requires the same amount of information as the  exact knowledge of $\psi$, in principle, however $f$ can be built on the basis of mathematical necessary conditions and physically well founded empirical considerations\cite{jpa1,jpa2}. Most important $f$ can be used for Monte Carlo sampling of electron configurations in space \cite{prb,jcpluca} to calculate $I_{nloc}[\rho]$ as $\int\rho({\bf r}) G\left(\rho({\bf r})\right)$. The empirical expression of $f$ employed is an exponential parametric form (or a simple variation of it when the spin of each particle is explicitly considered \cite{jcpluca}):
\begin{equation}
f({\bf r}_{2},....{\bf r}_{N}|{\bf r}_{1})=\Pi_{n=2,N}e^{E({\bf r}_{1})-\gamma
  V_{ee}({\bf r}_{1},{\bf r}_{n})}\times \Pi_{i>j\neq 1} e^{-\gamma V_{ee}({\bf
    r}_{i},{\bf r}_{j})}
\label{f1}
\end{equation}
with 
\begin{equation}
e^{-E({\bf r}_{1})}=\int\Pi_{n=2,N}e^{-\gamma
  V_{ee}({\bf r}_{1},{\bf r}_{n})}\times \Pi_{i>j\neq 1} e^{-\gamma V_{ee}({\bf
    r}_{i},{\bf r}_{j})} d{\bf r}_{2}...d{\bf r}_{N}.
\label{fnorm}
\end{equation}
Here $\gamma$ is a free parameter that is determined by a minimization procedure within the framework of the Levy-Lieb principle of DFT \cite{levy,lieb}.
With this set up, a Monte Carlo calculation for a gas of interacting electrons led to the conclusion: 
\begin{equation}
I_{nloc}[\rho]=\int \rho [A+B\log\rho]d{\bf r}
\label{inonl}
\end{equation}
Later on this expression was improved by adding explicitly the particles' spin so that the correct limit for non interacting particles was obtained (Thomas-Fermi term). The functional form remains the same except a refinement of the constant $A$ and $B$ and the addition of the Thomas-Fermi term. I have shown that $I_{nloc}[\rho]$ must correspond to a correlation energy \cite{jstat} once the Thomas-Fermi term is subtracted. This conclusion is fully consistent with the results of Sierraalta and Ludena, who have shown that for a gas of non interacting electrons the non local part of the kinetic energy corresponds to the Thomas-Fermi term of kinetic energy \cite{ludena}.
The expression $\rho\log\rho$ does not possess the correct scaling behaviour under coordinate scaling, ${\bf r}\to\lambda{\bf r}$, that is a very relevant property for the physical consistency of any energy functional\cite{levypedr}. Thus I suggested that the numerical result obtained expresses the first term of a Taylor expansion of $e^{wS[\rho]}$ (which instead has a consistent scaling). However, the validity of the numerical results have been put under discussion and my hypothesis of a link between the Shannon entropy and the kinetic functional has been refuted by a work of Trickey {\it et al.} \cite{trickey}.
Actually the conclusion of Ref.\cite{trickey} is that the Shannon Entropy cannot play a role in the description of the (kinetic) correlation energy. I personally do not agree with the conclusions of Ref.\cite{trickey}; 
however several aspects underlined there are extremely useful to narrow down the idea of a combination of numerical results and theoretical hypothesis (these aspects will be explicitly discussed later on). I can anticipate that the main point will be about which questions to address in order to definitively refute or prove the connection between the Shannon entropy and the correlation term of the energy functional.
Before setting the discussion mentioned above, there is the need of employing a more accurate numerical method, without a high degree of empiricism, to numerically determine the correlation part of the kinetic and of the potential energy for a uniform gas of interacting electron. If, using accurate methods, $T_{c}[\rho]$ and $V_{c}[\rho]$ are close to the expression of the Shannon entropy, then it is legitimate to ask the question whether or not this is a lucky case (as Trickey and coworkers claim) or if there is something more behind. If one is convinced of the second hypothesis, the key point is about finding a convincing interpretation of the correlation energy in the light of the meaning of Shannon entropy discussed before, that is a measure of correlation strength. 
In this perspective, a convincing argument is represented by the {\it Collins Conjecture}\cite{collins} in which the correlation energy of many-electron systems is written as a term proportional to the Shannon entropy. The connections between our results and the {\it Collins Conjecture} will be treated in detail in the discussion section. 
Below I report numerical results for $T_{c}[\rho]$ and $V_{c}[\rho]$ for a uniform gas of interacting electrons calculated with an advanced Quantum Monte Carlo method. Indeed, within a certain (extended) range of densities, relevant for condensed matter, the behaviour of $T_{c}[\rho]$ and $V_{c}[\rho]$ is that of $S[\rho]$ (each with the proper sign). In the subsequent section these results are discussed.

\section*{\sffamily \Large NUMERICAL RESULTS}
In this section I report the results for the kinetic and potential energy density per particle of a Quantum Monte Carlo study of a uniform electron gas of interacting electrons. Here the electronic correlations are described with very high accuracy, being the method used, Reptation Quantum Monte Carlo, state of the art among Quantum Monte Carlo approaches regarding electronic correlations \cite{markus}. The technical details of the simulation are reported in Ref.\cite{prlmarkus} here I will add only those details required for the current discussion.
The correlation energy as a function(al) of the density $\rho$ is defined as:
\begin{equation}
E_{c}[\rho]=E_{tot}[\rho]-T_{id}[\rho]-V_{ee}^{HF}[\rho]
\label{Ecorrqmc}
\end{equation}
where $T_{id}[\rho]$ is the kinetic energy of an ideal gas of non interacting electron at density $\rho$, which corresponds to the Thomas Fermi kinetic energy, and $V_{ee}^{HF}[\rho]$ is the electron-electron potential energy at density $\rho$ calculated taking as a trial wavefunction a Slater determinant of free particle orbitals (plane waves). According to the definition of Eq.\ref{ecorrqmc} the following definitions follow:
\begin{equation}
T_{c}[\rho]=T[\rho]-T_{id}[\rho]
\label{Tcorrqmc}
\end{equation}
\begin{equation}
V_{c}[\rho]=V_{ee}[\rho]-V_{ee}^{HF}[\rho].
\label{Vcorrqmc}
\end{equation}
and accordingly:
\begin{equation}
e_{c}(\rho)=\frac{1}{N}E_{c}[\rho]
\label{ecorrqmc}
\end{equation}
so that $E_{c}[\rho]=\int \rho [e_{c}(\rho)]d{\bf r}$
\begin{equation}
t_{c}(\rho)=\frac{1}{N}T_{c}[\rho]
\label{tcorrqmc}
\end{equation}
so that $T_{c}[\rho]=\int \rho [t_{c}(\rho)]d{\bf r}$
\begin{equation}
v_{c}(\rho)=\frac{1}{N}V_{c}[\rho]
\label{vcorrqmc}
\end{equation}
so that $V_{c}[\rho]=\int \rho [v_{c}(\rho)]d{\bf r}$.
The density range spans from 0.002 to 2.0 $\frac{e}{bohr^{3}}$ which corresponds, in the more familiar language of the Wigner-Seitz radius, $r_{s}$, to a range of 2.0-0.5 $bohr$. While the case of density $\rho=2.0 \frac{e}{bohr^{3}}$ shall be considered a high density regime, the other densities considered are those of interest for condensed matter systems under standard conditions. Such densities are contained in the intermediate density regime where general features of correlation functionals are likely to be more relevant for applications. Fig.\ref{tcorrfig} and Fig.\ref{vcorrfig} show that the behaviour of both the kinetic energy density (per particle) matches very closely that of a logarithm behaviour in the range from 0.002 to 0.25 $\frac{e}{bohr^{3}}$. Instead for higher values (2.0 $\frac{e}{bohr^{3}}$) the behaviour of the two quantities diverges from that of the logarithm obtained from the other densities. However, Fig.\ref{etotfig} shows that the total correlation energy density follows the logarithm behaviour. These data are not particularly surprising and are consistent with Quantum Monte Carlo data of the past employed to parametrize a largely used energy correlation functional of DFT \cite{cepald}. However in the context of this discussion these results provide a numerical evidence, within a certain range of relevant densities, of a $\log\rho$ behavior of $t_{c}$ and $v_{c}$; and for higher density it is their sum $e_{c}$ to behave as $\log\rho$. As a matter of fact, these results do not reject the idea of a link between Shannon entropy and electronic correlation functional of DFT. Actually these results motivate the search for a possible more profound link between the two concepts; below we discuss the possible aspects concerning the search of a more profound link.
\section*{\sffamily \Large DISCUSSION}
In the light of the discussion carried out in the previous sections here I will discuss some critical aspects about the possibility of a link between the correlation energy and Shannon entropy. The Quantum Monte Carlo results are certainly encouraging but their range of validity shall be properly considered. Trickey and coworkers \cite{trickey} have reported a limitation common to all numerical approaches based on sampling particle configurations, that is, the low density case. At low densities numerical convergence is hard to reach and accuracy is highly questionable, thus the extrapolation to low densities of a $\log$ behaviour for $t_{c}$, based on data at intermediate densities, leads to an unavoidable positivity violation, and thus to negative kinetic energy, which is a physical contradiction. More in general the problem of low accuracy applies also to $v_{c}$ and $e_{c}$. At high densities the problem of accuracy is instead minimized but the physics of the system becomes different since relativistic effects become increasingly more relevant.
These considerations imply a major restriction, that is, one shall consider the validity of the conclusions of a numerical study only in the range of densities considered in the calculations. A second important restriction concerns the fact that the data were obtained for a uniform gas of electrons where the density is constant in space. This is a strong simplification and, for example, the case of uniform density would not be consistent with the interpretation of Romera and Dehesa of Shannon Entropy, previously reported, as a descriptor of the ``delocalization'' of $\rho({\bf r})$ and thus of electronic correlations. At this point it becomes a matter of personal taste whether to believe in a more profound link between the two concepts or to believe that the results are an (un)fortunate coincidence.
In general, on the basis of the ground-breaking success of the concept of Shannon Entropy in a wide range of disciplines, I tend to believe in a more profound link between the two concepts, however one needs to be practical with specific questions and a possible working plan at a concrete level. Below I motivate my positive attitude and formulate a series of questions whose answer, in my view, would be relevant to give solid basis to this field of research.
\subsection*{\sffamily \large Monte Carlo Sampling as an encoding process and its correspondence with DFT via Shannon Entropy}
One point of crucial importance in this discussion is the following:
given any system of electrons at density $\rho({\bf r})$, it is always possible to define {\it a posteriori} its Shannon Entropy, $S=-\int[\rho({\bf r})\log\rho({\bf r})]d{\bf r}$; this is possible also for non interacting electrons, which have no correlations included. The actual key point of the discussion above is that the results of the QMC study do not concern a generic measure of correlations made {\it a posteriori}, but the functional form of correlation energies of interacting electrons. The functional form of such energies is a solution of the full quantum problem when the energy is written in terms of $\rho({\bf r})$, thus the similarity between this term and the Shannon entropy is not imposed, but emerges naturally as a solution of the full quantum problem. The first question to ask is whether or not it is a mere coincidence that methods which are based on statistical sampling/integration of electron configurations find a similar functional behaviour for the correlation energies in terms of $\rho$. The numerical accuracy corresponding to each specific method may be very different, however it cannot be denied that within the range of densities considered in the numerical calculations the functional form is the same\cite{nota}. As a matter of fact the sampling/averaging procedure of the correlation energy expressed in terms of $\rho({\bf r})$ is an encoding process of a set of $3N$-dimensional data into a set of $3$-dimensional data; this observation suggests to attempt a formulation of the analogy between correlation energies and Shannon entropy in terms of a process of encoding/decoding data.
In fact in the full quantum problem we have a $3N$-dimensional wavefunction which exactly expresses all the correlations between electrons. However, in the numerical study, when we reduce the expression of the energy to a $3$-dimensional quantity, the result is that the correlation energy is expressed in form of (proportional to) a quantity which in information theory is a measure of the uncertainty hidden in its $3$-dimensional distribution (due to an encoding process). In the language of information theory $S$ quantifies the average amount of information (but does not provide the information itself) needed to express the realization of a certain event.\\
If we use this analogy, can we then interpret $\int\rho({\bf r})[t_{c}({\bf r})]d{\bf r}$, $\int\rho({\bf r})[v_{c}({\bf r})]d{\bf r}$ and $\int\rho({\bf r})[e_{c}({\bf r})]d{\bf r}$ (with the appropriate sign) as quantities whose {\it leading} term ($\int\rho\log\rho$) is a sort of index of the average quantity of information needed to express the explicit/exact $3N$-dimensional data?\\
At this stage this interpretation is certainly speculative but at the same time very appealing. In fact the Hohenberg-Kohn theorem of DFT \cite{hk} sets a one-to-one correspondence between the $3N$-dimensional wavefunction of ground state $\psi({\bf r}_{1}....{\bf r}_{N})$ and the $3$-dimensional electron density $\rho({\bf r})$.
As a matter of fact the passage from $\psi$ to $\rho$ implies a process of encoding/integration (and vice versa the process from $\rho$ to $\psi$ corresponds to a decoding process) and the revolutionary essence of the theorem is that in $\rho({\bf r})$ are contained (coded) all the properties of the ground state, even its wavefunction (see e.g.\cite{ijqc,chapbook}).
Following the arguments above it becomes natural the following question: according to the idea of encoding/decoding of information theory, does the $\log\rho$ form of  $t_{c}$, $v_{c}$ and $e_{c}$ tell us that the correlation terms of the universal functional of Hohenberg and Kohn expresses the fact that the correlation energy corresponds (is proportional) to the average quantity of information needed to explicitly express the exact many-body behavior of the electrons?\\
In order to be more concrete on this point I will propose the possibility of an analogy with the horse racing example illustrated by Petz in his book \cite{petz}.
\subsection*{\sffamily \large Horse Racing Analogy}
If we want to communicate the name of the winning horse in a horse racing where all horses have the same probability to win (uniform distribution) the minimum length of a message in binary code is determined by the Shannon entropy.
For example if the race is made by $8$ horses then we have: $S=\sum_{i=1,8}\frac{1}{8}\log_{2}\frac{1}{8}=3$; this means that I can identify each horse in an exact way with a message expressed by a combination of $1$'s and $0$'s of length $3$ ($3$ bits). Here the {\it ``event''} is the victory of one specific horse; the probability distribution is about which horse can win and it is very important to notice that the probability does not say anything about the characteristics which identify each horse and distinguish it from the others (e.g. the name, color, age etc etc); such characteristics  are communicated via the encoding process(a string of $0$'s and $1$'s). The idea can be extended in a straightforward way to a non uniform distribution, in which case the events are not all equally probable and the Shannon entropy expresses the {\it average length of the message}, that is the average number of bits needed to communicate any of the possible outcome of the racing.
In the case of electrons, $\rho({\bf r})$ expresses the event of finding one electron at a given point in space ${\bf r}$ due to the (average) action of the other electrons which is implicit in the shape of $\rho({\bf r})$. This implies that $\rho({\bf r})$ expresses the probability of realization of specific electronic configurations (in a $3$-dimensional space, neglecting the spin for simplicity). Being normalized to $N$ it then expresses the fact that electrons are indistinguishable and thus this event would be true independently of the choice of an electron of reference. As in the example of the horses, $\rho({\bf r})$ tells us about the likelihood of an event (a certain electronic conformation in space), but does not tell us about the explicit characteristics of the event, that is the action of all the other electrons on the electron of reference  which led to such an event.
In order to express the explicit action of all the other electrons one needs to specify two-body, three-body....$N$-body correlations corresponding to the event (something equivalent to a string of bits as the string required to communicate the winning horse in the horse racing example).
In fact, for a system of non interacting electrons the probability of finding one electron in space does not depend on the action of the other electrons, thus the one-body information contained in ${\bf r}_{1}\equiv{\bf r}$ is sufficient and the encoding of information related to $({\bf r}_{2},.....{\bf r}_{N})$ not required. In this perspective, if we consider the electronic correlations per particle as the overall action of all electrons on one specific electron, then the correlation energy, in the language of information theory, is nothing else than the average amount of information (i.e. of action) needed to explicitly express electronic configurations. Non interacting electrons do not make any action, thus they do not have a correlation energy, instead for interacting electrons the action is taken into account by the encoding process, that is the sampling/averaging on the $3N$-dimensional space. The example discussed here is, at this stage, only a possible suggestive interpretation and it is not supported by any further mathematical/formal argument. However an encouraging support may be found in the work of D.M.Collins\cite{collins} and in its related developments. The similarity with the idea of Collins is discussed in the next paragraph.
\subsection*{\sffamily \large Collins Conjecture}
In a seminal paper, D.M.Collins\cite{collins} put forward the following conjecture:
\begin{equation}
E_{c}=\xi\sum_{j}n_{j}\log n_{j}
\label{collins1}
\end{equation}
where $\xi$ is a proper constant and $n_{j}$ the occupation number of the $j$-th state/spin-orbital. Later on, Ziesche\cite{ziesche} extended this concept to the momentum distribution, $\rho({\bf k})$. He explicitly discussed the case of a uniform electron gas and identified the Shannon entropy with the correlation energy. The Shannon entropy in this case is written as:
\begin{equation}
S=-\int_{0}^{\infty}d\left(\frac{{\bf k}}{{\bf k}_{F}}\right)\rho({\bf k})\log \rho({\bf k})
\label{zies}
\end{equation}
with ${\bf k}_{F}$ the normalizing Fermi momentum at the given electron density.
He concludes that:\\
{``\it $s$ measures, at least for the uniform electron gas, the correlation strength''}.\\
However, as in my case, on the basis of some numerical results,  Ziesche cannot confirm Collin's conjecture for low densities. Anyway, few years later, Collins conjecture was numerically proven to be valid for a series of small molecules, by Ramirez {\it et al}\cite{soriano}. They employ a sufficiently accurate quantum chemical approach based on configuration interaction wavefunctions and go beyond the case of a uniform gas of electrons. These results are rather encouraging regarding the validity of Collins conjecture at least in first approximation, that is the leading term of $E_{c}$ is proportional to the Shannon entropy as defined by Collins and Ziesche, additional terms are required for the very low and very high density. In this perspective, the numerical results that I have presented here  and my hypothesis on $t_{c}$, $v_{c}$ and $e_{c}$ may be considered an extension of Collins conjecture to the case of $S$ written in terms of $\rho({\bf r})$ at least in the range of electron densities employed in the calculations. However, all the arguments given in the two preceeding paragraphs need a critical appraisal in order to give a credibility and/or solidity to the discussion.
\subsection*{\sffamily \large Warning}
The Shannon Entropy is defined as the, $-\int\rho[\log\rho] d{\bf r}$, and, for example $t_{c}$ is positive by definition, thus one should define the proper prefactor and understand its meaning in order to have consistency between the meaning of $S$ and that of each term of the correlation energy.
Moreover, I must clarify that I am not proposing, $\log\rho$, as a universal energy density for the Hohenberg-Kohn functional; as Ziesche\cite{ziesche} also underlines, the low density case of a uniform gas would contradict my conclusions. My message is that we have numerical evidence for a gas of electrons, in  a well specified (and relevant) range of densities, that the {\it the leading} term (at least) of the correlation energy density has the form of $\log\rho$. I then make the suggestion that it may exist a universal functional whose explicitly form involves in some way the expression of, $\log\rho$, and thus it can be related to the concept of Shannon entropy.
The correlation energy functional is known analytically in the limit of high densities and in the limit of low densities (see e.g.\cite{prbfo}), thus test limiting cases are known and this would help in the construction of a functional whose leading term is the $\log\rho$ (at least for a gas of electrons at least at intermediate densities). This is fully consistent with other results found in literature and based on the Collins conjecture\cite{ziesche, soriano}.
In DFT, after the initial enthusiasm, the theoretical development has not evolved as initially hoped and expected. The actual development in the field is going towards the use of an increasing amount of elaborated/elegant empiricism sold as conceptual development but as a matter of fact justified by an encouraging but yet not sufficient success in numerical applications. New ideas are needed, and of course they are very likely to be associated with high risk of failure; the ideas expressed in this paper enters in such a category.
\subsection*{\sffamily \large A Sketch of a possible Research Roadmap}
The idea of investigating the concept of sampling/integrating as an encoding process and relate it to the concept of Shannon entropy is certainly appealing, but in order to make a concrete step towards this idea one should first have  at least numerical evidence that the link between the correlation energies of QMC and the Shannon entropy is not a coincidence. A useful suggestion would be that of performing Quantum Monte Carlo calculations of $t_{c}$, $v_{c}$ and $e_{c}$ for a representative series of atoms or simple molecules and check how close is their functional form to $\log\rho({\bf r})$. If the dominant term is still the $\log\rho({\bf r})$, then one would be motivated to proceed towards further investigation. An encouraging results is certainly that of Ref.\cite{soriano}, but it is still not sufficient.
Next, an effort should be done in finding a general functional form of $\log\rho({\bf r})$ which has the correct formal behavior (e.g. coordinate scaling) and leads to the correct high density and low density limit.
If this can be done successfully, then the process of encoding a set of $3N$-dimensional data in a set of $3$-dimensional data in DFT can be viewed from the perspective of information theory. It may turn out that the tools of information theory are implicit concepts in the statement of the Hohenberg-Kohn theorem of DFT.
\section*{\sffamily \Large CONCLUSIONS}
The idea of Shannon Entropy has already been employed in well founded applications in the physics of many-electron systems; in particular regarding electronic correlations. In this paper I have reported numerical data which, within a certain range of validity, encourage the idea of a connection between electronic correlation energies and Shannon Entropy. I have speculated that, if one accepts a general validity of the numerical result beyond their current limitations, it may be possible to interpret the numerical results as a process of encoding data according to the procedures of information theory. The Quantum Monte Carlo results that I have shown suggest an extension of Collins conjecture to the case of $\rho({\bf r})$. 
I suggested numerical studies which may enforce or definitively refute the connection discussed above. Certainly, in practice, it will be needed a major effort in terms of numerical investments; however, more delicate may be the question of convincing a sufficiently large portion of researchers to be pioneers in this field. \redc{Implicitly the idea of encoding/decoding data is already used in quantum chemistry and electronic structure calculations; the subject of ``{\it Inverse Chemistry}'' (see e.g.\cite{markusrei}) is gaining popularity and, in my view, the path of ``decoding'' many-electron properties from $\rho({\bf r})$ using Shannon Entropy would be very useful to the inverse problem. Finally, the concept of ``electronic correlations'' of DFT or quantum chemistry may be no more sufficient for the accuracy required by modern studies of chemistry and material physics; it does not exists an operator to define this concept and thus electronic correlations cannot be directly observed. Most probably the concept of ``Entanglement'' would be more powerful in this sense (see e.g. Ref.\cite{kais}). Eisert, Cramer and Plenio \cite{eisert} treat the concept of entropy of entanglement as ``{\it quite profound quantity}'' and discuss it in terms of locality of interactions and correlation functions of quantum systems. I have already discussed, implicitly, some work which refers to entropy of entanglement, but the additional point here is to emphasize that key properties such as the ``area laws of entanglement entropy'', discussed by Eisert, Cramer and Plenio, may play an important role in the future developments of electronic structure approaches (in general and of DFT in particular). In conclusion, the concept of {\it ``Information''} was considered by  John Wheeler to be at the basis of the very fundamental laws of physics \cite{wheeler} , thus of electronic correlations, as this paper attempts to emphasize.} 

\subsection*{\sffamily \large ACKNOWLEDGMENTS}
I thank Carlo Pierleoni and Markus Holzmann for allowing me to use the QMC data and for a critical reading of the manuscript. I thank Luca Ghiringhelli for reading the paper and put forward several suggestions. This work was supported by the Deutsche Forschungsgemeinschaft (DFG) within the Heisenberg Program (grant code DE 1140/5-1).

\clearpage




\clearpage

\begin{figure}
\caption{\label{tcorrfig} Kinetic correlation energy per particle as a function of the electron density. Vertical lines confine the {\it linear regime}, where the energy is proportional to $\log\rho$.}
\end{figure}

\begin{figure}
\caption{\label{vcorrfig} Coulomb electron-electron correlation energy per particle as a function of the electron density. Vertical lines confine the {\it linear regime}, where the energy is proportional to $\log\rho$.}
\end{figure}

\begin{figure}
\caption{\label{etotfig} Total correlation energy per particle as a function of the electron density. For densities of the order of 2.0 $\frac{e}{bohr^{3}}$ kinetic and Coulomb correlation term do not follow the $\log$ behaviour indicated by data at smaller densities, however their sum, and thus the total correlation energy follows still a $\log$ behavior.}
\end{figure}


\clearpage

\begin{center}
\includegraphics[width=0.95\columnwidth,angle=0,keepaspectratio=true]{kcorr-tot.eps}
\end{center}
\vspace{0.25in}
\hspace*{3in}
{\Large
\begin{minipage}[t]{3in}
\baselineskip = .5\baselineskip
Figure 1 \\
Luigi Delle Site\\
Int. J.\ Quant.\ Chem.
\end{minipage}
}
\begin{center}
\includegraphics[width=0.95\columnwidth,angle=0,keepaspectratio=true]{u_corr-tot.eps}
\end{center}
\vspace{0.25in}
\hspace*{3in}
{\Large
\begin{minipage}[t]{3in}
\baselineskip = .5\baselineskip
Figure 2 \\
Luigi Delle Site\\
Int. J.\ Quant.\ Chem.
\end{minipage}
}
\begin{center}
\includegraphics[width=0.8\columnwidth,angle=-90,keepaspectratio=true]{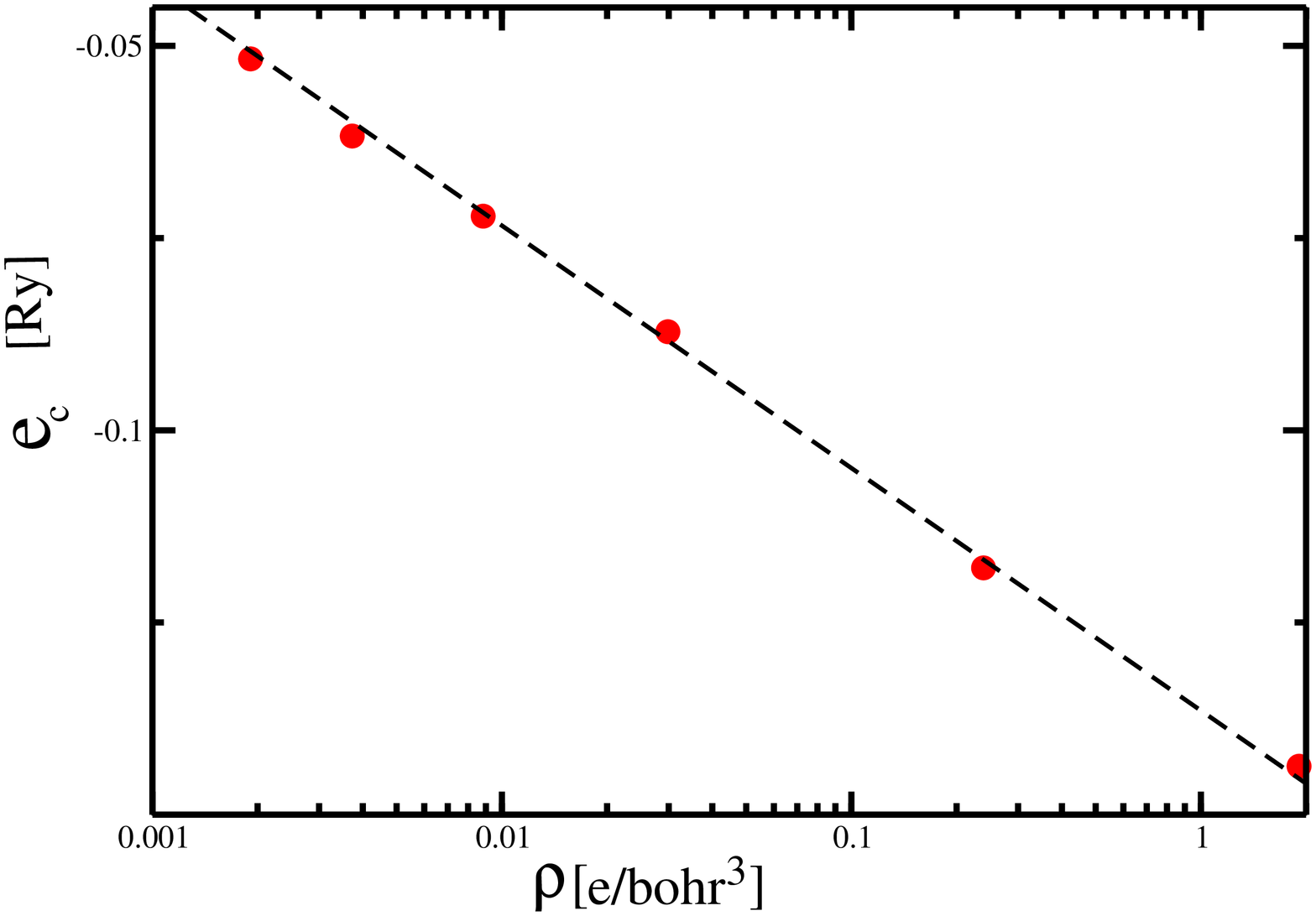}
\end{center}
\vspace{0.25in}
\hspace*{3in}
{\Large
\begin{minipage}[t]{3in}
\baselineskip = .5\baselineskip
Figure 3 \\
Luigi Delle Site\\
Int. J.\ Quant.\ Chem.
\end{minipage}
}

\end{document}